\documentclass[peerreview]{IEEEtran}
\IEEEoverridecommandlockouts

\usepackage{cite}
\usepackage{amsmath,amssymb,amsfonts}
\usepackage{algorithmic}
\usepackage{graphicx}
\usepackage{textcomp}
\usepackage{xcolor}
\newcommand{\parbreak}{ \vspace{0.4 em}}
\def\BibTeX{{\rm B\kern-.05em{\sc i\kern-.025em b}\kern-.08em
    T\kern-.1667em\lower.7ex\hbox{E}\kern-.125emX}}

\usepackage{amsthm}

\usepackage{physics}

\renewcommand{\trace}[1]{\mathrm{Tr}\left\{#1\right\}}

\newcommand{\Aset}{\mathcal{A}}

\newcommand{\Dset}{\mathcal{D}}

\newcommand{\Hset}{\mathcal{H}}

\newcommand{\Lset}{\mathcal{L}}

\newcommand{\Uset}{\mathcal{U}}
\newcommand{\Vset}{\mathcal{V}}
\newcommand{\Qset}{\mathcal{Q}}

\newcommand{\Xset}{\mathcal{X}}
\newcommand{\Yset}{\mathcal{Y}}
\newcommand{\Zset}{\mathcal{Z}}

\newcommand{\hm}{\hat{m}}
\newcommand{\hM}{\widehat{M}}
\newcommand{\channel}{\mathcal{N}}
\newcommand{\eps}{\varepsilon}
\newcommand{\opC}{\mathcal{C}}

\newcommand{\prob}[1]{\Pr\left(#1\right)}

\usepackage{mathrsfs}

\usepackage{bbm}
\newcommand{\identity}{\mathbbm{1}}

\theoremstyle{remark}	\newtheorem{theorem}{Theorem}
\theoremstyle{remark}	\newtheorem{lemma}[theorem]{Lemma}
\theoremstyle{remark}	
\theoremstyle{remark}	
\theoremstyle{remark} \newtheorem{definition}{Definition}
\theoremstyle{remark} \newtheorem{remark}{Remark}
\theoremstyle{remark}

\begin{document}

\title{The Interference Channel with Entangled Transmitters
\thanks{The authors acknowledge the financial support by the Federal Ministry of Education and Research of Germany (BMBF) in the programme of “Souverän. Digital. Vernetzt.”. Joint project 6G-life, project identification number: 16KISK263 (JH, AM, HA, CD). Furthermore, the work was supported by the following grants of the BMBF 16KISQ028 (CD, JH), 16KIS2196 (CD, JH), 16KISQ169 (CD, AM) and 16KIS2234 (CD, HA).}
}

\author{\IEEEauthorblockN{Jonas Hawellek, Athin Mohan, Hadi Aghaee and Christian Deppe}

\IEEEauthorblockA{\textit{Institute for Communications Technology} \\
\textit{Technische Universität Braunschweig}\\
Braunschweig, Germany \\
\{jonas.hawellek, athin.mohan, hadi.aghaee, christian.deppe\}@tu-braunschweig.de}
}

\maketitle

\begin{abstract}
This paper explores communication over a two-sender, two-receiver classical interference channel, enhanced by the availability of entanglement resources between transmitters. The central contributions are an inner and outer bound on the capacity region for a general interference channel with entangled transmitters. It addresses the persistent challenge of the lack of a general capacity formula, even in the purely classical case, and highlights the striking similarities in achievable rate expressions when assessing quantum advantages. Through a concrete example, it is shown that entanglement can significantly boost performance in certain types of channels.
\end{abstract}


\setcounter{page}{1}

\section{Introduction}

\parbreak In the context of 6G research, there is a growing vision to enhance classical networks by integrating a quantum layer \cite{Granelli_Bassoli_Nötzel_Fitzek_Boche_daFonseca_Guerrieri_2022}. This quantum layer is designed to improve communication performance by enabling higher data rates, enhanced security, and lower latency. Early studies have already shown notable improvements in specific scenarios. However, much of the research has focused on point-to-point connections \cite{Ekert_1991}\cite{Bennett_Shor_Smolin_Thapliyal_1999} or smaller, simpler networks such as the MAC \cite{Hsieh_Devetak_Winter_2008, Leditzky_Alhejji_Levin_Smith_2020, Nötzel_2020, Seshadri_Leditzky_Siddhu_Smith_2023, Pereg_Deppe_Boche_2023} or broadcast channels \cite{Yard_Hayden_Devetak_2011}\cite{Pereg_Deppe_Boche_2021}. In contrast, this work addresses the interference channel, a more complex model involving two senders and two receivers.

\parbreak The interference channel (IC) is a key model in network information theory. Previous strategies involving additional resources, such as transmitter cooperation, have demonstrated potential for improving communication rates \cite{Prabhakaran_Viswanath_2011}. This raises the question of whether shared entanglement between transmitters alone - while keeping the rest of the model fully classical - can also provide a meaningful advantage in communication rates.

\parbreak In \cite{Quek_Shor_2017}, Quek and Shor demonstrated the advantages of using entangled transmitters by analyzing the separation of rates.
They presented a specific instance of an IC based on the CHSH game \cite{Clauser_Horne_Shimony_Holt_1969}, but primarily focused on the analysis of superquantum non-local correlations derived from the PR-box model proposed by Popescu and Rohrlich \cite{Popescu_Rohrlich_1994}.
Later, Leditzky et al. \cite{Leditzky_Alhejji_Levin_Smith_2020} explored a related scenario involving multiple access channels (MAC) and concluded that any channel model derived from non-local games \cite{Brassard_Broadbent_Tapp_2005} exhibits a capacity advantage when entanglement is shared among transmitters.
Non-local games, including the CHSH game, are special cooperative games in which two distant parties agree on a joint strategy to maximize their probability of winning, without communicating during gameplay.
By leveraging shared entanglement, they can enhance the probability of winning beyond what any classical strategy, absent of communication, can achieve \cite{Brassard_Broadbent_Tapp_2005}.
Concurrently, Nötzel \cite{Nötzel_2020} examined MACs in which non-zero communication rates can only be achieved with the assistance of entanglement.
Recently, further analysis on the MAC with entangled transmitters was conducted in \cite{Pereg_Deppe_Boche_2023}, providing bounds on the capacity region.

\parbreak In this paper, we explore the effect of entanglement shared among the transmitters in the IC.

\parbreak The structure of this paper is as follows:
In Section \ref{sec:definitions}, we provide basic definitions and introduce the system model.
Section \ref{sec:classicalic} discusses bounds on the capacity of the classical IC, as established in the literature.
Section \ref{sec:mainResultsGeneral} presents our primary findings regarding the classical IC with entangled transmitters.
An illustrative example is discussed in Section \ref{sec:example}.
Sections \ref{sec:proofHK} to \ref{sec:proofCardinality} contain the proofs related to our findings.
Finally, Section \ref{sec:summaryDiscussion} offers a summary and discussion of our results and outlines potential directions for further research.

\section{Definitions}
\label{sec:definitions}

\subsection{Basic Definitions}
The state of a quantum system $A$ is described by a density operator $\rho$ on the Hilbert space $\mathcal{H}_A$. The set $\Dset(\mathcal{H}_A)$ is the set of all such density operators.
A positive operator-valued measure (POVM) is defined as a collection of positive semi-definite measurement operators $\Lset = \{L_x\}_{x\in\Xset}$ that satisfy the condition $\sum_{x\in\Xset}L_x = \identity$ with $\identity$ being the identity operator on the respective Hilbert space.
The probability of obtaining the measurement outcome $x$ from the state $\rho$ is given by $p_X(x)=\trace{L_x\rho}$.
A quantum channel $\channel_{A \rightarrow B}$ is a completely positive and trace-preserving map from $\Dset(\mathcal{H}_A)$ to $\Dset(\mathcal{H}_B)$.
A measurement channel is defined as a quantum-to-classical channel $\mathcal{L}_{A \to X}$ mapping $\rho\mapsto\mathcal{L}_{A \to X}(\rho)=\sum_{x\in\Xset}\Tr{L_x\rho}\ketbra{x}$. In this expression, $\{L_x\}_{x\in\Xset}$ is a POVM, and the states $\ket{x}$, where $x\in\Xset$, denote the classical output symbols.
A discrete memoryless interference channel $(\Xset_1,\Xset_2,P_{Y_1,Y_2|X_1,X_2},\Yset_1,\Yset_2)$ with two inputs $X_k \in \Xset_k$, $k\in\{1,2\}$, and two outputs $Y_k \in \Yset_k$, $k\in\{1,2\}$, is described by a collection of conditional probability mass functions $P_{Y_1,Y_2|X_1,X_2}(y_1,y_2|x_1,x_2)$ on $\Yset_1 \times \Yset_2$.

\subsection{Interference Channel with Entangled Transmitters}
We consider a classical IC $P_{Y_1, Y_2|X_1,X_2}$ with entanglement shared between the transmitters.
The entanglement resources of transmitter 1 and transmitter 2 are denoted by $E_1$ and $E_2$, respectively, and the corresponding Hilbert space is given by $\Hset_{E_1 E_2}\equiv \Hset_{E_1}\otimes \Hset_{E_2}$.

\begin{definition}
    \label{def:codeICwE}
    A $(2^{nR_1},2^{nR_2},n)$ code for the classical IC $P_{Y_1,Y_2|X_1,X_2}$ with
    entangled transmitters consists of the following components:
    \begin{itemize}
    \item An entangled state $\varphi_{E_1 E_2}\in\Dset(\Hset_{E_1 E_2})$ shared between the transmitters;
    \item Two message sets $[1:2^{nR_1}]$ and $ [1:2^{nR_2}]$, assuming $2^{nR_k}$ is an integer;
    \item A pair of encoding POVMs $\Lset_1^{(m_1)}=\{ L^{(m_1)}_{x_1^n} \}_{x_1^n\in\mathcal{X}_1^n}$ and 
    $\Lset_2^{(m_2)}=\{ L^{(m_2)}_{x_2^n} \}_{x_2^n\in\Xset_2^n}$ on $E_1$ and $E_2$, respectively;
    \item Two decoding functions $g_1:\Yset_1^n\to [1:2^{nR_1}]$ and ${g_2:\Yset_2^n\to [1:2^{nR_2}]}$.
    \end{itemize}
    We denote this code by $\mathscr{C}=(\varphi,\Lset_1,\Lset_2,g_1, g_2)$.
\end{definition}

\begin{figure}[t]
\centerline{\includegraphics[width=0.75\textwidth]{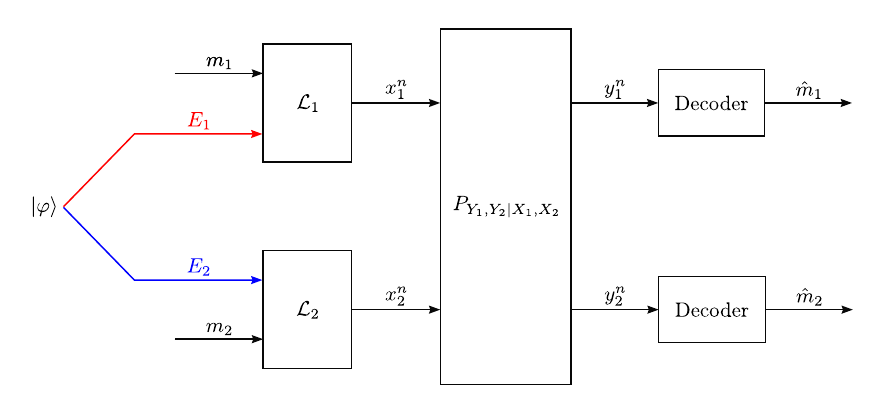}}
\caption{The classical interference channel $P_{Y_1,Y_2|X_1,X_2}$ with entanglement resources (quantum systems) shared between the transmitters.
The entanglement resources of transmitter 1 and transmitter 2 are marked in red and blue, respectively.}
\label{fig:ICentangledTx}
\end{figure}

The corresponding communication scheme is illustrated in Figure~\ref{fig:ICentangledTx}.

The entangled pair $E_1 E_2$ is pre-shared between the transmitters.
Alice $k$ now selects a message $m_k$ from her message set $[1:2^{nR_k}]$.
To encode this message, Alice $k$ performs the measurement $\Lset_k^{(m_k)}$ on her share $E_k$ of the entangled pair.
This measurement yields the classical outcome $x_k^n \in \mathcal{X}_k^n$, which is transmitted through $n$ uses of the classical IC $P_{Y_1,Y_2|X_1,X_2}$.
The joint input distribution to the channel is thus given by
\begin{align}%
f(x_1^n,x_2^n|m_1,m_2)
= \trace{\left(L^{(m_1)}_{x_1^n}\otimes L^{(m_2)}_{x_2^n} \right) \varphi_{E_1 E_2}} \,.
\end{align}
After the action of the channel on the inputs, Bob $k$ receives the channel output $y_k^n$ and estimates the message as ${\hm_k=g_k(y_k^n)}$.

The conditional probability of error at receiver $k$ of the code $\mathscr{C}=(\varphi,\Lset_1,\Lset_2,g_1, g_2)$ is
\begin{equation}
P_{e,k}^{(n)}(\mathscr{C}|m_1,m_2)= %
\sum_{y_k^n: g_k(y_k^n)\neq m_k} \left[ \sum_{(x_1^n,x_2^n)\in \Xset_1^n\times \Xset_2^n} f(x_1^n,x_2^n|m_1,m_2) P_{Y_1,Y_2|X_1,X_2}^n(y_1^n, y_2^n|x_1^n,x_2^n) \right] %
\end{equation}
for $(m_1,m_2)\in [1:2^{nR_1}]\times [1:2^{nR_2}]$.
The message-average probability of error at receiver $k$ is defined as
\begin{equation}
\overline{P}_{e,k}^{(n)}(\mathscr{C})\equiv %
\frac{1}{2^{n(R_1+R_2)}} \sum_{m_1=1}^{2^{nR_1}} \sum_{m_2=1}^{2^{nR_2}}    P_{e,k}^{(n)}(\mathscr{C}|m_1,m_2) \,.
\label{eq:Message_Average_Error}
\end{equation}

A rate pair $(R_1,R_2)$ is considered achievable with entangled transmitters if, for every $\eps>0$ and sufficiently large $n$, there exists a $(2^{nR_1},2^{nR_2},n)$ code such that the message-average probability of error satisfies $\overline{P}_{e,k}^{(n)}(\mathscr{C})\leq\eps $. 
The capacity region $\opC_{\text{ET}}(P_{Y_1,Y_2|X_1,X_2})$ of the classical IC with entangled transmitters
is defined as the closure of the set of achievable rate pairs $(R_1,R_2)$, where the subscript `ET' denotes the presence of  entanglement resources between the transmitters. %

\section{Bounds on the Capacity of the Classical Interference Channel}
\label{sec:classicalic}

The classical interference channel has been extensively studied \cite{ElGamal_Kim_2011}, yet its general capacity region remains unknown.
In scenarios with strong and very strong interference conditions, we possess exact capacity expressions.
However, for the general case, we must rely on both inner and outer bounds to manage the uncertainty surrounding the capacity region.

\subsection{Han-Kobayashi Inner Bound}
\label{susbec:clasicalInner}
The best-known inner bound to the capacity region was established by Han and Kobayashi \cite{Han_Kobayashi_1981} and later simplified by Chong et al. \cite{Chong_Motani_Garg_ElGamal_2008}.
This inner bound is outlined in Lemma~\ref{lem:HKbound}, following an introduction to its main components.

Consider a discrete memoryless IC $P_{Y_1,Y_2|X_1,X_2}$ and let $\mathcal{R}_{\text{HK}}$ be the set of rates $(R_1,R_2)$ that satisfy \begin{align}
    R_1 &< I(X_1;Y_1|V_0,U_2) \label{eq:cHK_ineq1}\\
    R_2 &< I(X_2;Y_2|V_0,U_1) \\
    R_1+R_2 &< I(X_1;Y_1|V_0,U_1,U_2)+I(U_1,X_2;Y_2|V_0) \\
    R_1+R_2 &< I(X_1,U_2;Y_1|V_0,U_1)+I(U_1,X_2;Y_2|V_0,U_2) \\
    R_1+R_2 &< I(X_1,U_2;Y_1|V_0)+I(X_2;Y_2|V_0,U_1,U_2) \\
    2R_1+R_2 &< I(X_1,U_2;Y_1|V_0)+I(X_1;Y_1|V_0,U_1,U_2) \nonumber \\
    &\quad+I(U_1,X_2;Y_2|V_0,U_2) \\
    R_1+2R_2 &< I(X_1,U_2;Y_1|V_0,U_1)+I(U_1,X_2;Y_2|V_0) \nonumber \\
    &\quad+I(X_2;Y_2|V_0,U_1,U_2) \label{eq:cHK_ineq7}
\end{align}
for some probability mass function
\begin{equation*}
    p_{V_0,X_1,U_1,X_2,U_2}(v_0,x_1,u_1,x_2,u_2) = p_{V_0}(v_0) p_{X_1,U_1|V_0}(x_1,u_1|v_0) p_{X_2,U_2|V_0}(x_2,u_2|v_0)
\end{equation*}
defined at the transmitters.
The auxiliary time-sharing variable $V_0$ can be bounded by $|V_0|\leq7$. For the other auxiliary variables arising from superposition coding, we have $|U_1|\leq|X_1|+4$ and $|U_2|\leq|X_2|+4$.

\begin{lemma}
    \label{lem:HKbound}
    Any rate pair $(R_1,R_2) \in \mathcal{R}_{\text{HK}}$ is achievable for the discrete memoryless IC $P_{Y_1,Y_2|X_1,X_2}$ \cite[Theorem 6.4]{ElGamal_Kim_2011}.
    Hence, the capacity region $\opC$ of the general discrete memoryless IC satisfies
    \begin{equation}
        \opC \supseteq \mathcal{R}_{\text{HK}} \,.
    \end{equation}
\end{lemma}

\begin{remark}
    While this rate region encompasses most of channels and reduces to the actual capacity region in the case of strong interference conditions, it is known to be strictly smaller than the general capacity region.
    There exist interference channels that are known to fall outside of this rate region \cite{Nair_Xia_Yazdanpanah_2015}.
\end{remark}

\subsection{Outer Bound}
\label{subsec:classicalOuter}

An outer bound on the general capacity region for an IC $P_{Y_1,Y_2|X_1,X_2}$ is provided by the following Lemma \ref{lem:upperbound} \cite{ElGamal_Kim_2011}\cite{Sato_1977}.

\begin{lemma}
    \label{lem:upperbound}
    Let $\mathcal{R}_{\text{o}}$ be the set of rate pairs $(R_1,R_2)$ that satisfy
    \begin{align}
        R_1 &\leq I(X_1;Y_1|V_0,X_2) \label{eq:cOuter_ineq1}\\
        R_2 &\leq I(X_2;Y_2|V_0,X_1) \\
        R_1 + R_2 &\leq I(X_1,X_2;Y_1,Y_2|V_0) \,, \label{eq:cOuter_ineq3}
    \end{align}
    where $V_0$ is an auxiliary time-sharing variable.
    The capacity region $\opC$ of the general discrete memoryless IC satisfies
    \begin{equation}
        \opC \subseteq \mathcal{R}_{\text{o}} \,.
    \end{equation}
\end{lemma}

\section{Main Results}
\label{sec:mainResultsGeneral}

\subsection{Inner Bound}
\label{susbsec:inner}

In this section, we provide an inner bound on the general capacity region for the classical two-transmitter, two-receiver IC $P_{Y_1,Y_2|X_1,X_2}$ with transmitters sharing entanglement.
Our approach is inspired by the Han-Kobayashi inner bound for the capacity region without entanglement assistance, as presented in Section \ref{susbec:clasicalInner}.
Given that this is currently the best-known achievable rate region, we anticipate being able to achieve at least the same rates by employing similar coding techniques while also leveraging the entanglement shared between transmitters.
To illustrate the potential advantage of entanglement assistance, we aim to demonstrate that the entanglement-assisted capacity region is larger than the unassisted classical one.
If this can be established, it would facilitate the identification of specific instances where entanglement provides a measurable advantage.
Without entanglement assistance, only a limited number of channels are known to fall outside the Han-Kobayashi rate region, as evidenced in examples such as those presented in \cite{Nair_Xia_Yazdanpanah_2015}.
For these few examples, we will certainly not be able to ascertain from our results whether entanglement provides an additional advantage.

\parbreak To define the rate region, we perform rate-splitting and superposition coding.
Let $m_k$, $k\in\{1,2\}$, be split into two independent parts $m_k=(m'_k,m''_k)$, with $R_k'$ and $R_k''$ being the respective rates such that $R_k=R_k'+R_k''$.
While $m_k'$ is intended to be decoded by both receivers, $m_k''$ is decoded solely by the respective receiver $k$, allowing that receiver to utilize the information obtained from both $m_1'$ and $m_2'$.
The message $m_k'$ is encoded into a codeword $u_k(m_k'|v_0)$, and, in the context of superposition coding, $m_k''$ is encoded into the codeword $v_k(m_k',m_k''|v_0)$, conditioned on some common random variable $V_0$ that is accessible to all transmitters and receivers.
Based on both codewords $u_k$ and $v_k$, the POVM $\mathcal{L}^{(v_0, v_k, u_k)}_k$ is selected, which will produce the final encoded channel input $x_k$.

\parbreak Let $\mathcal{P}$ be the set of probability mass functions $p_{V_0,V_1,U_1,V_2,U_2,X_1,X_2,Y_1,Y_2}$ that factor as
\begin{multline}
p_{V_0,V_1,U_1,V_2,U_2,X_1,X_2,Y_1,Y_2}(v_0,v_1,u_1,v_2,u_2,x_1,x_2,y_1,y_2)
= p_{V_0}(v_0) p_{V_1,U_1|V_0}(v_1,u_1|v_0) p_{V_2,U_2|V_0}(v_2,u_2|v_0) \\
\qquad\cdot\trace{\left( L_1(x_1|v_0,v_1,u_1)\otimes L_2(x_2|v_0,v_2,u_2) \right)\varphi_{E_1 E_2}}
\cdot P_{Y_1,Y_2|X_1,X_2}(y_1,y_2|x_1,x_2) \,.
\label{eq:Distribution_HK_inner}
\end{multline}
for any input distributions $p_{V_0}p_{V_1,U_1|V_0}p_{V_2,U_2|V_0}$, shared quantum states $\varphi_{E_1 E_2}$ and encoding POVMs $\Lset_1\otimes\Lset_2$.
For any such probability mass function $p \in \mathcal{P}$ and, consequently, the associated fixed input distribution, shared quantum state, and encoding POVMs, let $\mathcal{R}_{\text{ET-HK}}(P_{Y_1,Y_2|X_1,X_2},p)$ be the set of rates $(R_1,R_2)$ satisfying
\begin{align}
    R_1 &< I(V_1,U_1;Y_1|V_0,U_2) \label{eq:HK_ineq1} \\
    R_2 &< I(V_2,U_2;Y_2|V_0,U_1) \\
    R_1+R_2 &< I(V_1;Y_1|V_0,U_1,U_2)+I(U_1,V_2,U_2;Y_2|V_0) \\
    R_1+R_2 &< I(V_1,U_2;Y_1|V_0,U_1)+I(U_1,V_2;Y_2|V_0,U_2) \\
    R_1+R_2 &< I(V_1,U_1,U_2;Y_1|V_0)+I(V_2;Y_2|V_0,U_1,U_2) \\
    2R_1+R_2 &< I(V_1,U_1,U_2;Y_1|V_0)+I(V_1;Y_1|V_0,U_1,U_2) + I(U_1,V_2;Y_2|V_0,U_2) \\
    R_1+2R_2 &< I(V_1,U_2;Y_1|V_0,U_1)+I(U_1,V_2,U_2;Y_2|V_0) + I(V_2;Y_2|V_0,U_1,U_2)  \,. \label{eq:HK_ineq7}
\end{align}
The inequalities \eqref{eq:HK_ineq1} -- \eqref{eq:HK_ineq7} are similar to the classical results represented in \eqref{eq:cHK_ineq1} -- \eqref{eq:cHK_ineq7}.
In our setting, however, it is not possible to simplify these inequalities further by rewriting the codeword $V_k$ resulting from superposition coding with the channel input $X_k$.
As previously noted, the encoding POVMs are selected depending on $V_k$, and their outputs are $X_k$, the channel inputs.
Because these POVMs operate on the entangled pairs shared between the transmitters, they may introduce correlations between the channel inputs $X_1$ and $X_2$, which are intended to exploit advanatges by the entanglement.
Consequently, the data processing inequality is inapplicable on $V_k$ and $X_k$.
Therefore, we cannot bound and simplify the expressions above by replacing $V_k$ with $X_k$.
Our main result is now presented in the following theorem.

\begin{theorem}
\label{thm:InnerBound}
Consider a discrete memoryless IC $P_{Y_1,Y_2|X_1,X_2}$ with entangled transmitters.
We define an achievable rate region $\mathcal{R}_{\text{ET-HK}}(P_{Y_1,Y_2|X_1,X_2})$ as follows:
\begin{equation}
    \mathcal{R}_{\text{ET-HK}}(P_{Y_1,Y_2|X_1,X_2})=
    \bigcup_{\substack{ p \in \mathcal{P} }}
    \mathcal{R}_{\text{ET-HK}}(P_{Y_1,Y_2|X_1,X_2}, p) \,.
    \label{eq:Etx-HK-region}
\end{equation}
This rate region consists of any rate pairs $(R_1, R_2)$ that satisfy the conditions in \eqref{eq:HK_ineq1} -- \eqref{eq:HK_ineq7} for some $p\in\mathcal{P}$.
The proof of this theorem is given in Section~\ref{sec:proofHK}.

\parbreak Since any achievable rate region is a subset of the capacity region, the capacity region $\opC_{\text{ET}}(P_{Y_1,Y_2|X_1,X_2})$ of the IC with entangled transmitters satisfies
\begin{equation}
    \opC_{\text{ET}}(P_{Y_1,Y_2|X_1,X_2}) \supseteq \mathcal{R}_{\text{ET-HK}}(P_{Y_1,Y_2|X_1,X_2}) \,.
\end{equation}
\end{theorem}

\begin{lemma}
    \label{lem:purification}
    It is sufficient to consider pure states $\varphi_{E_1E_2}\equiv\ketbra{\phi_{E_1E_2}}$ to exhaust the union in \eqref{eq:Etx-HK-region}.
    The proof of this assertion follows as in \cite{Pereg_Deppe_Boche_2024} and is provided in the appendix.
\end{lemma}

\begin{lemma}
    The union in \eqref{eq:Etx-HK-region} is achieved with $|\Vset_0|\leq7$.
    The proof of this statement is provided in Section~\ref{sec:proofCardinality}.
\end{lemma}

\begin{remark}
    The dimensions of the entangled state are not bounded.
    Consequently, our rate region assumes unlimited entanglement assistance.
    Furthermore, we give no bounds on the alphabet sizes of the auxiliary variables $\Uset_1$ and $\Uset_2$.
    Due to the shared entanglement utilized in the encoding process, $X_1$ and $X_2$ cannot be regarded as statistically independent.
    As a result, straightforward applications of the Fenchel-Eggleston-Carathéodory theorem \cite{Eggleston_1966} lead to mutual dependencies in the alphabet sizes at both receivers.
    Nevertheless, the authors remain confident that finite bounds exist.
    The reasoning behind this confidence is that the rate region bears similarities to the classical case, which has established bounds (cf. Section \ref{susbec:clasicalInner}).
    Furthermore, similar considerations in \cite{Pereg_Deppe_Boche_2024} established finite bounds in the MAC case, although the structure of its rate region is simpler.
\end{remark}

\begin{remark}
    The encoding POVMs produce a single classical codeword each.
    For the selection of POVMs, we utilize both $u_k(m_k')$ and $v_k(m_k', m_k'')$, which arises from superposition coding.
    The union over any encoding POVMs $\Lset_1 \otimes \Lset_2$ yields the same rate region when conditioning solely on the messages $m_k=(m_k',m_k'')$.
    Consequently, while superposition coding is not essential for achievability, it facilitates the definition of two messages, of which one is employed in both decoders, without requiring further specification of the POVMs.
\end{remark}

We have previously noted that the rate region resembles the classical results presented in Section \ref{susbec:clasicalInner}.
The primary distinction lies in the additional union over quantum states and encoding POVMs.
Consequently, it is difficult to discern any advantage from entanglement merely by comparing the rate regions; an example is necessary to illustrate this advantage.
Furthermore, we discussed that the Han-Kobayashi rate region serves only as an inner bound on the classical capacity.
A larger classical achievable region may be found, and there are known examples of channels that lie outside this region \cite{Nair_Xia_Yazdanpanah_2015}.
The same holds true for our rate region, which further emphasizes the necessity of providing an example.

\subsection{Outer Bound}
\label{subsec:outer}

We state an outer bound on the capacity region $ \mathcal{C}_{ET}$ of the general IC with entangled transmitters.
Let $\mathcal{P}'$ be the set of probability mass functions $p'_{V_0,V_1,V_2,X_1,X_2,Y_1,Y_2}$ that factor as
\begin{multline}
p'_{V_0,V_1,V_2,X_1,X_2,Y_1,Y_2}(v_0,v_1,v_2,x_1,x_2,y_1,y_2) \\
= p_{V_0}(v_0)p_{V_1,V_2|V_0}(v_1,v_2|v_0)\cdot\trace{\left( L_1(x_1|v_0,v_1)\otimes L_2(x_2|v_0,v_2) \right)\varphi_{E_1 E_2}}\cdot P_{Y_1,Y_2|X_1,X_2}(y_1,y_2|x_1,x_2) \,.
\label{eq:Distribution_outer}
\end{multline}
for any input distributions $p_{V_0,V_1,V_2,X_1,X_2,Y_1,Y_2}$, shared quantum states $\varphi_{E_1 E_2}$ and encoding POVMs $\Lset_1\otimes\Lset_2$.
For any such probability mass function $p' \in \mathcal{P}'$ and the associated fixed input distribution, shared quantum state, and encoding POVMs, let $\mathcal{R}_{\text{ET-o}}(P_{Y_1,Y_2|X_1,X_2},p')$ be the set of rates $(R_1,R_2)$ satisfying
\begin{align}
    R_1 &< I(V_1;Y_1|V_0,V_2) \label{eq:outer_ineq1} \\
    R_2 &< I(V_2;Y_2|V_0,V_1) \\
    R_1+R_2 &< I(V_1,V_2;Y_1,Y_2|V_0)  \,. \label{eq:outer_ineq3}
\end{align}
In comparison to the inner bound, $p'$ does not assume statistical independence between $V_1$ and $V_2$ given $V_0$.
Furthermore, we do not utilize a rate-splitting strategy where only part of each message is available to the other receiver.
Instead, we assume that the receivers have access to the entire message from the other transmitter while decoding the intended message.
The sum rate is bounded by the combined mutual information from any channel input to any output, irrespective of the information availability at the encoders and decoders.

\parbreak We state the outer bound in the following theorem.

\begin{theorem}
    \label{thm:OuterBound}

Consider a discrete memoryless IC $P_{Y_1,Y_2|X_1,X_2}$ with entangled transmitters.
We define the rate region $\mathcal{R}_{\text{ET-o}}(P_{Y_1,Y_2|X_1,X_2})$ as follows:
\begin{equation}
    \mathcal{R}_{\text{ET-o}}(P_{Y_1,Y_2|X_1,X_2})=
    \bigcup_{\substack{ p' \in \mathcal{P}' }}
    \mathcal{R}_{\text{ET-o}}(P_{Y_1,Y_2|X_1,X_2}, p') \,.
    \label{eq:Etx-outer-region}
\end{equation}
This rate region includes any rate pairs $(R_1, R_2)$ that satisfy the conditions in \eqref{eq:outer_ineq1} -- \eqref{eq:outer_ineq3} for some $p'\in\mathcal{P}'$.

\parbreak The capacity region $\opC_{\text{ET}}(P_{Y_1,Y_2|X_1,X_2})$ of the IC with entangled transmitters satisfies
\begin{equation}
    \opC_{\text{ET}}(P_{Y_1,Y_2|X_1,X_2}) \subseteq \mathcal{R}_{\text{ET-o}}(P_{Y_1,Y_2|X_1,X_2}) \,.
\end{equation}
    
\end{theorem}

The proof of Theorem~\ref{thm:OuterBound} is provided in Section~\ref{sec:proofOuter}.
Lemma~\ref{lem:purification} is also applicable to the outer bound, such that it is sufficient to consider pure states $\varphi_{E_1 E_2}$. The proof is given in the appendix, where $U_1=U_2\equiv\emptyset$.

\begin{lemma}
    The union in \eqref{eq:Etx-outer-region} is achieved with $|\Vset_0|\leq3$.
    The proof follows from standard application of the Fenchel-Eggleston-Carathéodory Theorem \cite{Eggleston_1966}, analogous to the proof presented in Section~\ref{sec:proofCardinality} for the inner bound.
\end{lemma}

Similar to the inner bound, the outer bound resembles the classical result presented in Section~\ref{subsec:classicalOuter}.
The primary distinction lies in the union over shared quantum states and encoding POVMs.
Additionally, as with the inner bound, it is not possible to equate $V_k$ with $X_k$ for $k\in\{1,2\}$ based on the same reasoning.
This makes it unclear whether an advantage from the shared entanglement can exist based on comparing the outer bounds.
Furthermore, the outer bounds may not be achievable under any strategy.

\begin{remark}
    In the classical case, it is possible to establish capacity regions in the cases of strong or very strong interference \cite{Costa_Gamal_1987}\cite{ElGamal_Kim_2011}.
    To achieve this, the (very) strong interference conditions are utilized in the converse to obtain tight upper bounds.
    Similar to the general upper bound discussed here, the expressions do not directly consider the channel inputs $X_k$, $k\in\{1,2\}$.
    Rather, they consider $V_k$ to which the conditions cannot be applied.
    Consequently, while the achievability proofs can be successfully established in these cases using similar techniques, the converse arguments do not hold, and we therefore omitted these cases entirely.
\end{remark}

\section{Channel Example}
\label{sec:example}
We consider an example of an IC in which entanglement shared between transmitters enhances communication rates.
Looking at the MAC, we know from \cite{Leditzky_Alhejji_Levin_Smith_2020} and \cite{Pereg_Deppe_Boche_2023} that entanglement shared between the transmitters provides an advantage for the class of channels defined from non-local games.
They specifically focused on the magic square game due to the existence of a perfect quantum strategy that simplifies the analysis.
In the following, we will construct a similar yet more general channel model of an IC based on the magic square game.

\subsection{Magic Square Game}
\label{subsec:magicsquaregame}

The magic square game is a cooperative game with two players and a referee.
The central component is a $3\times3$ square in which the players can assign binary values to each cell.
The gameplay is as follows:
The referee selects one out of the nine cells at random.
Suppose that the selected cell is $(r,c)$.
Player 1 is informed on the row index $r$ and Player 2 is informed on the column index $c$.
Their task now is to fill in the corresponding row or column with three binary values.
The game is won if three constraints are met:
\begin{itemize}
    \item the binary values given for the overlapping cell $(r,c)$ coincide,
    \item row $r$ has even parity, and
    \item column $c$ has odd parity.
\end{itemize}
Since it is a cooperative game, the shared goal of the players is to fulfill these requirements, which means in particular that they want to maximize the probability of agreeing on the value in the overlapping cell.
To achieve this, the players can agree on a strategy before the game starts, but communication at a later time is prohibited.
If the players are limited to classical strategies, it is impossible for them to always win the game.
An illustrative deterministic approach to win the game is shown in Table~\ref{Table:Magic_Classical}.
\begin{table}
\renewcommand{\arraystretch}{1.3}
\caption{Magic square game: deterministic strategy}
\label{Table:Magic_Classical}
\centering
\begin{tabular}{| c| c| c| }
\hline
 0 & 0 & 0 \\ 
\hline
 0 & 1 & 1 \\  
\hline
 1 & 0 & ?    \\
\hline
\end{tabular}
\end{table}
For any pair $(r,c)$ of row and column indices selected by the referee, the players fill in the corresponding row and column according to this table.
This ensures that they agree on the overlapping cell, except when $(r,c)=(3,3)$.
It is impossible to fill in this cell, as the parity constraints require different values in that particular case.
Consequently, the probability of winning the game is $\frac{8}{9}$, assuming $(r,c)$ is drawn uniformly at random.
Furthermore, it is shown in \cite{Brassard_Broadbent_Tapp_2005} that this probability is the maximum achievable among any classical strategies, not limited to deterministic ones.

\parbreak If the players share an entanglement resource, however, it becomes possible for them to win the game with probability~1.
Prior to the game, prepare the state
\begin{equation}
    \label{eq:Magic_A1A2}
    \ket{\phi_{A_1' A_1'' A_2' A_2''}} =\frac{1}{2}\bigl(
        \ket{00} \ket{11} + \ket{11} \ket{00}  - \ket{01} \ket{10} - \ket{10} \ket{01}
    \bigl) \,,
\end{equation}
and send $A_k',A_k''$ to player k.
After having received the row and column indices, the players play the game according to the quantum strategy outlined in Table~\ref{Table:Magic_Quantum}.
\begin{table}
\renewcommand{\arraystretch}{1.3}
\caption{Magic square game: quantum strategy}
\label{Table:Magic_Quantum}
\centering
\begin{tabular}{| c| c| c| }
\hline
 $\mathsf{X}\otimes\identity$ & $\mathsf{X}\otimes \mathsf{X}$ & $\identity\otimes \mathsf{X}$ \\ 
\hline
 $-\mathsf{X}\otimes \mathsf{Z}$ & $\mathsf{Y}\otimes \mathsf{Y}$ & $-\mathsf{Z}\otimes \mathsf{X}$ \\  
\hline
 $\identity\otimes \mathsf{Z}$ & $\mathsf{Z}\otimes \mathsf{Z}$ & $\mathsf{Z}\otimes \identity$    \\
\hline
\end{tabular}
\end{table}
The table assigns an observable to each cell, where $\mathsf{X}$, $\mathsf{Y}$, and $\mathsf{Z}$ are the Pauli operators.
Player 1 simultaneously measures the three observables specified in the row indexed by $r$ on the joint system $A_1'A_1''$, while Player 2 measures the observables in column $c$ on the system $A_2'A_2''$.
The measurement outcomes of each observable are then recorded in the corresponding cells.
By following this strategy, the players are guaranteed to agree on the overlapping cell with probability~1 \cite{Brassard_Broadbent_Tapp_2005}.

\subsection{Channel Model and Capacity Results}

Leditzky et al. \cite{Leditzky_Alhejji_Levin_Smith_2020} defined a MAC based on the magic square game, where the channel is noiseless for input combinations that win the game.
Otherwise, the channel selects purely random outputs.
In this manner, cooperation between the transmitters that assists in winning the game also results in an increase in communication rates.
This definition of a MAC can be readily transformed into an IC with the same properties.
The precise definition of this channel is provided below.

\parbreak In the definition of the magic square game previously discussed in Section \ref{subsec:magicsquaregame}, the referee selects two questions $q_1\in\mathcal{Q}_1 = \{1,2,3\}$ and $q_2\in\mathcal{Q}_2 = \{1,2,3\}$, which were previously referred to as the row and column indices, uniformly at random and hands them to the players.
We now consider the questions to be messages that the players wish to communicate over an IC.
Therefore, these questions are no longer provided by the referee.
Rather, they are already available at the transmitters.
The players select their respective binary answer sequences $a_1\in\mathcal{A}_1=\{ 0,1 \}^3$ and $a_2\in\mathcal{A}_2=\{ 0,1 \}^3$ based on the implemented strategies $f_k:\mathcal{Q}_k\to \mathcal{A}_k$, for $k=1,2$, such as the quantum strategy outlined in Table \ref{Table:Magic_Quantum}.
The function $f_k$ thus serves as our encoder.
In this setup, the channel effectively takes on the role of the referee.
Since it does not select the questions, the transmitters input both the questions $q_k$ and answers $a_k$ to the channel.
The game is won if $(q_1,q_2,a_1,a_2)\in\mathscr{G}$, where $\mathscr{G}$ is the winning set defined as
\begin{align*}
   \mathscr{G} =\big\{\big( q_1&,q_2,a_1,a_2 \big) \in \Qset_1\times\Qset_2\times\Aset_1\times\Aset_2: \\
        & a_1[q_2]=a_2[q_1], \\
        & a_1[1]+ a_1[2] + a_1[3] \mod 2=0, \\
        & a_2[1]+ a_2[2] + a_2[3] \mod 2=1
        \big\} \;.
\end{align*}
This simply reflects the rules of the magic square game.

\parbreak We define a classical IC $P_{Y_1,Y_2|X_1,X_2}$ with
\begin{align}
    \Xset_k &=\mathcal{Q}_k\times \mathcal{A}_k \,,\; k=1,2\\
    \Yset_k&= \mathcal{Q}_k \,,\; k=1,2
\end{align}
such that
\begin{equation}
    P_{Y_1,Y_2|X_1,X_2}\big( \hat{q}_1,\hat{q}_2 \,\big|  q_1,a_1,q_2,a_2  \big) \\
    =
    \begin{cases}
        \delta_{q_1, \hat{q}_1} \delta_{q_2, \hat{q}_2} &\text{if } (q_1,q_2,a_1,a_2)\in\mathscr{G}\,,\\
        \frac{1}{|\mathcal{Q}_1| |\mathcal{Q}_2|} &\text{otherwise.}
    \end{cases}
\label{Equation:MS_channel}
\end{equation}
The referee evaluates the channel inputs $X_1=(q_1,a_1)$ and $X_2=(q_2,a_2)$, and precisely outputs the questions to the decoders if the game is won, meaning $Y_1=q_1$ and $Y_2=q_2$ with probability~1.
Otherwise, if the game is lost, the channel outputs are chosen uniformly at random from the question sets.

\parbreak We now analyze the rates achievable with both the classical and the quantum strategy, demonstrating that the latter can achieve higher rates.
Seshadri et al. \cite{Seshadri_Leditzky_Siddhu_Smith_2023} established a bound on the sum-rate of a MAC based on a non-local game when any classical strategy is employed.
However, their approach to this bound is not readily applicable to the IC.
To address this, we introduce a simple upper bound for the IC by effectively treating it as a MAC.
Thus, the same sum-rate bound that is known from \cite{Seshadri_Leditzky_Siddhu_Smith_2023} and \cite{Pereg_Deppe_Boche_2023} can also be applied to the IC.

\parbreak Define $Y=(Y_1, Y_2)$.
This results in a MAC $P_{Y|X_1,X_2}$ based on the magic square game, for which the sum-rate is bounded by $R_1 + R_2 \leq 3.02$ for any classical strategy \cite{Seshadri_Leditzky_Siddhu_Smith_2023}.
We know that the capacity region $\opC_{\text{MAC}}$ of a MAC is given by the set of rate pairs $(R_1,R_2)$ satisfying
\begin{align}
    R_1 &\leq I(X_1,Y|V_0,X_2) \\
    R_2 &\leq I(X_2,Y|V_0,X_1) \\
    R_1 + R_2 &\leq I(X_1,X_2;Y|V_0) \,,
\end{align}
where $V_0$ is an auxiliary time-sharing variable \cite[Theorem~4.3]{ElGamal_Kim_2011}.
This serves as an upper bound to the outer bound $\mathcal{R}_{\text{o}}$ in Lemma \ref{lem:upperbound} on the capacity region of the IC, since we provide more information to the receivers.
This observation was previously made by Sato \cite{Sato_1977}, who argued that the decoder at the MAC is not limited to one of the two channel outputs, unlike the situation with the IC.
Therefore, we have
\begin{equation}
    \opC \subseteq \mathcal{R}_{\text{o}} \subseteq \opC_{\text{MAC}} \,,
\end{equation}
indicating that an upper bound on the sum-rate of the MAC is also applicable to the underlying IC.
In particular, the capacity region of the discussed IC based on the magic square game, with only classical common randomness shared between the transmitters, therefore satisfies
\begin{equation}
\mathcal{C}(P_{Y_1,Y_2|X_1,X_2})\subseteq \left\{
 (R_1,R_2) \,:\;
	R_1 + R_2 \leq 3.02
\right\} \,.
\label{eq:classicalstrategybound}
\end{equation}
Since this is a very straightforward evaluation, a tighter bound may indeed be found for the IC.
By splitting the channel output $Y$ of the MAC representation into two separate outputs $Y_1$ and $Y_2$ and considering a scenario without cooperation between the receivers, there is less information available to each of the decoders.
Consequently, we expect a reduction in performance.

\parbreak Since entanglement between the transmitters allows them to win the the game with probability~1 by employing the quantum strategy, the sum rate $R_1 + R_2 = 2\log(3) \approx 3.17$ is achievable with entangled transmitters.

\parbreak As as result of the previous discussion, we state the following theorem.
\begin{theorem}
    The capacity region of the IC based on the magic square game with entangled transmitters is strictly larger than the capacity region when the transmitters are restricted to classical correlations only.
    In the classical case, the sum-rate satisfies $R_1+R_2\leq3.02$, while in the entanglement-assisted case, we have $R_1+R_2\leq2\log_2(3)\approx3.17$ and $R_1+R_2=2\log_2(3)$ is achievable.
\end{theorem}

The main difference between the example presented here and the MAC example in \cite{Pereg_Deppe_Boche_2023} is that the channel outputs of the IC remain separate.
We noted that the MAC bound also applies to the IC, as the game-based IC can be considered a generalization of its MAC version.

\parbreak Considering the game-based IC with cooperating receivers, we can simplify it to a single receiver, essentially obtaining a MAC version of the game.
More generally, we can analyze any IC with arbitrarily strong receiver cooperation.
Each of these configurations will be subject to the same upper sum-rate bound as the corresponding MAC, which can be effectively treated as an IC with perfect receiver cooperation.
In the case of any channel based on a non-local game, a quantum strategy will consistently outperform any classical strategy.

\parbreak Receiver cooperation may be achieved through additional resources shared between the receivers or by physically combining the receivers into a single entity, as is the case in the MAC scenario.
Keeping in mind the goal of identifying further examples that demonstrate an advantage from entangled transmitters, there exists another approach to achieving an equivalent of receiver cooperation: if the channel model itself provides partial information about the second message to the receivers alongside the intended message.

\section{Proof of Theorem~\ref{thm:InnerBound} (Inner Bound)}
\label{sec:proofHK}

To define the rate region, we employ rate-splitting and superposition coding.
Let $m_k$, $k\in\{1,2\}$, be split into two independent parts, such that $m_k=(m'_k,m''_k)$, with $R_k'$ and $R_k''$ denoting the respective rates, thus yielding $R_k=R_k'+R_k''$.
While $m_k'$ is intended to be decoded by both receivers, $m_k''$ is decoded solely by the respective receiver $k$, allowing that receiver to utilize the information obtained from both $m_1'$ and $m_2'$.

\parbreak Suppose that the transmitters share $n$ copies of the entangled state
\begin{align}
\varphi_{E_1^n E_2^n}\equiv \varphi_{E_1 E_2}^{\otimes n} \,.
\end{align}
We use superposition coding to generate code books for the messages $m_1$ and $m_2$.
\begin{figure}[t]
\centerline{\includegraphics[width=0.5\textwidth]{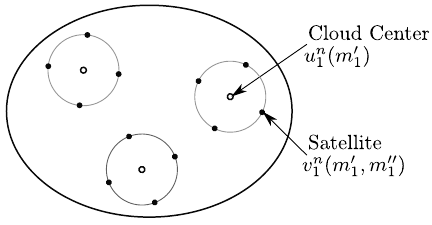}}
\caption{Superposition coding at transmitter 1.
Message $m_1'$ is intended to be decoded by both receivers and is encoded into distant cloud center codewords $u_1^n$.
Message $m_1''$ is decoded solely by receiver 1 after the corresponding cloud center ($m_1'$) has already been determined.}
\label{fig:SuperpositionCoding}
\end{figure}
An illustrative example is given in Figure~\ref{fig:SuperpositionCoding}.

\parbreak
\subsubsection{Code Construction}

Fix a probability mass function $p(v_0)p(u_1,v_1|v_0)p(u_2,v_2|v_0)$.
Generate a random i.i.d. sequence $v_0^n\sim \prod_{i=1}^{n}P_{v_0}(v_{0}[i])$.
For each transmitter $k$, $k\in\{1,2\}$, select $2^{nR^\prime_k}$ conditionally independent codewords $u^{n}_k (m^{\prime}_k)$, one for each $m^{\prime}_k \in [1,2^{nR^\prime_k}]$, according to the i.i.d. distribution $\prod_{i=1}^{n}P_{U_k|V_0}(u_{k}[i]|v_{0}[i])$.
For each $u^{n}_k (m^{\prime}_k)$, select $2^{nR^{\prime\prime}_k}$ conditionally independent codewords $v^{n}_k (m^{\prime}_k,m^{\prime\prime}_k )$, one for each $m^{\prime\prime}_k \in [1,2^{nR^{\prime\prime}_k}]$, accoording to the i.i.d. distribution $\prod_{i=1}^{n}P_{V_k|V_0, U_k}(v_{k}[i]|v_{0}[i], u_k[i])$. 
The auxiliary codebooks above are revealed to both the transmitters and the receivers.

\parbreak
\subsubsection{Encoder k}
Given $m_k=[m^{\prime}_k,m^{\prime\prime}_k]$ and the codebooks defined above, perform the measurement  $\bigotimes_{i=1}^n\left(
\Lset_k\big(v_0[i], v_{k}[i](m^{\prime}_k,m^{\prime\prime}_k),u_{k}[i](m^{\prime}_k)\big) \right)$ on the entangled system $A_k^n$, and transmit the measurement outcome $x_k^n$ through the channel.
This results in the following input distribution,
\begin{align}
f(x_1^n,x_2^n|m_1,m_2)&=  \trace{ \left( L_1^n\big(x_1^n \,\big| v_0^n, v_1^n(m^{\prime}_1,m^{\prime\prime}_1), u_1^n(m^{\prime}_1) \big) \otimes 
L_2^n\big(x_2^n \,\big| v_0^n, v_2^n(m^{\prime}_2,m^{\prime\prime}_2), u_2^n(m^{\prime}_2) \big)   \right)  
\varphi_{A_1^n A_2^n}
  } \,,
\end{align}
where we use the shorthand notation  $L_k^n\big( x_k^n \,\big| v_0^n, v_k^n, u_k^n\big)\equiv 
\bigotimes_{i=1}^n L_k \big( x_{k}[i] \,\big| v_0[i], v_{k}[i], u_{k}[i]\big)  $, 
for $k\in\{1,2\}$.

\parbreak
\subsubsection{Decoders}
Given $y^n_1$, decoder $1$ attempts to identify a unique message pair $(\hm^{\prime}_1,\hm^{\prime\prime}_1)$ and some $\hm^{\prime}_2$ such that $(v_0^n, v^{n}_1(\hat{m}^{\prime}_1,\hat{m}^{\prime\prime}_1), u^{n}_1(\hat{m}^{\prime}_1), u^{n}_2(\hat{m}^{\prime}_2), y^{n}_1) \in\Aset_\delta^{(n)}(p_{V_0, V_1, U_1, U_2, Y_1})$.
Similarly, given $y^n_2$, decoder $2$ attempts to identify a unique message pair $(\hm^{\prime}_2,\hm^{\prime\prime}_2)$ and some $\hm^{\prime}_1$ such that $(v_0^n, v^{n}_2(\hat{m}^{\prime}_2,\hat{m}^{\prime\prime}_2), u^{n}_1(\hat{m}^{\prime}_1), u^{n}_2(\hat{m}^{\prime}_2), y^{n}_2) \in\Aset_\delta^{(n)}(p_{V_0, V_2, U_1, U_2, Y_2})$.

\parbreak
\subsubsection{Analysis of Probability of Error}
We assume that the messages $(m^{\prime}_1,m^{\prime\prime}_1, m^{\prime}_2,m^{\prime\prime}_2)$ are chosen at random according to a uniform distribution.
By symmetry, we may assume, without loss of generality, that the transmitters send the messages $m^{\prime}_1 = 1, m^{\prime\prime}_1 = 1$ and $m^{\prime}_2 = 1, m^{\prime\prime}_2 = 1$.
First, consider decoder $1$, for which we have the following error events:
\begin{align}
\mathscr{E}_{10}=& \{  (V_0^n, V_1^n(1,1), U_1^n(1), U_2^n(1), Y_1^n) \notin \Aset_{\delta}^{(n)}(p_{V_0,V_1,U_1,U_2,Y_1})\} \\
\mathscr{E}_{11}=& \{  (V_0^n, V_1^n(1,m^{\prime\prime}_1), U_1^n(1), U_2^n(1), Y_1^n) \in \Aset_{\delta}^{(n)}(p_{V_0,V_1,U_1,U_2,Y_1})\text{, for some $m^{\prime\prime}_1\neq 1$  } \}\\
\mathscr{E}_{12}=& \{  (V_0^n, V_1^n(m^{\prime}_1,m^{\prime\prime}_1), U_1^n(m^{\prime}_1), U_2^n(1), Y_1^n)\in \Aset_{\delta}^{(n)}(p_{V_0,V_1,U_1,U_2,Y_1})\text{, for some $m^{\prime}_1\neq 1$, $m^{\prime\prime}_1$}  \}\\
\mathscr{E}_{13}=& \{  (V_0^n, V_1^n(1,m^{\prime\prime}_1), U_1^n(1), U_2^n(m^{\prime}_2), Y_1^n)\in \Aset_{\delta}^{(n)}(p_{V_0,V_1,U_1,U_2,Y_1})\text{, for some $m^{\prime}_2\neq 1$ and $m^{\prime\prime}_1\neq 1$} \}\\
\mathscr{E}_{14}=& \{  (V_0^n, V_1^n(m^{\prime}_1,m^{\prime\prime}_1), U_1^n(m^{\prime}_1), U_2^n(m^{\prime}_2), Y_1^n)\in \Aset_{\delta}^{(n)}(p_{V_0,V_1,U_1,U_2,Y_1})\text{, for some $m^{\prime}_1\neq 1$, $m^{\prime}_2\neq 1$, $m^{\prime\prime}_1$} \}
\end{align}

Hence, the expected probability of error for decoder $1$ is upper-bounded by

\begin{align}
\mathbb{E}\left[P_{e,1}^{(n)}(\mathscr{C}|1,1)\right] %
&\leq %
 \prob{ \mathscr{E}_{10} }% 
+ \prob{ \mathscr{E}_{11}|\mathscr{E}^c_{10}}+ \prob{ \mathscr{E}_{12}|\mathscr{E}^c_{10}}
+ \prob{ \mathscr{E}_{13}|\mathscr{E}^c_{10} }+ \prob{ \mathscr{E}_{14}|\mathscr{E}^c_{10} } \,.
\end{align}

We bound each term.
By the law of large numbers, $\prob{\mathscr{E}_{10}}$ tends to zero as $n\rightarrow\infty$.
Now observe that given $\mathscr{E}^c_{10}$, $(V_0^n, U_1^n(1), U_2^n(1)) \in \Aset_{\delta}^{(n)}(p_{V_0,U_1,U_2})$.
Given this, and by standard method-of-types arguments \cite{Csiszár_Körner_2011}\cite[Theorem 1.3]{Kramer_2008}, the remaining probabilities of error $\prob{ \mathscr{E}_{11}|\mathscr{E}^c_{10}}$, $\prob{ \mathscr{E}_{12}|\mathscr{E}^c_{10}}$, $\prob{ \mathscr{E}_{13}|\mathscr{E}^c_{10} }$, and $\prob{ \mathscr{E}_{14}|\mathscr{E}^c_{10} }$ tend to zero as $n\rightarrow\infty$, provided that
\begin{align}
    R_1'' &< I(V_1;Y_1|V_0,U_1,U_2) \\
    R_1' + R_1'' &< I(U_1,V_1;Y_1|V_0,U_2) \\
    R_1'' + R_2' &< I(V_1,U_2;Y_1|V_0,U_1) \\
    R_1' + R_1'' + R_2' &< I(U_1,V_1,U_2;Y_1|V_0) \,.
\end{align}
By employing a similar approach for decoder 2, we can further derive the bounds
\begin{align}
    R_2'' &< I(V_2;Y_2|V_0,U_1,U_2) \\
    R_2' + R_2'' &< I(U_2,V_2;Y_2|V_0,U_1) \\
    R_1' + R_2'' &< I(V_2,U_1;Y_2|V_0,U_2) \\
    R_1' + R_2' + R_2'' &< I(U_2,V_2,U_1;Y_2|V_0) \,.
\end{align}
We substitute $R^{\prime\prime}_k = R_k - R^{\prime}_k$ and add the constraints $0 \leq R^{\prime}_k \leq R_k$ for $k\in\{1,2\}$.
With these bounds, applying the Fourier-Motzkin elimination \cite{ElGamal_Kim_2011} results in a rate region equivalent to $\mathcal{R}_{\text{ET-HK}}(P_{Y_1,Y_2|X_1,X_2})$ as specified by inequalities \eqref{eq:cHK_ineq1} - \eqref{eq:cHK_ineq7}.
For any rate pair within that rate region, we conclude that the average probabilities of error $\mathbb{E}\left[\overline{P}_{e,1}^{(n)}(\mathscr{C})\right]$ and $\mathbb{E}\left[\overline{P}_{e,2}^{(n)}(\mathscr{C})\right]$ over the message sets tend to zero as $n\rightarrow\infty$.
Hence, there exists a $(2^{nR_1},2^{nR_2},n)$ code such that the message average error probability becomes arbitrarily small for sufficiently large $n$.
\qed

\section{Proof of Theorem~\ref{thm:OuterBound} (Outer Bound)}
\label{sec:proofOuter}

We prove the outer bound on the capacity region $\mathcal{C}_{EA}$ for a discrete memoryless IC with entangled transmitters.
Each message $m_k$ at the receiver is assigned a codeword $a_k$, which is then input into the encoder that utilizes the shared entanglement resource to produce the codeword $x_k$.
Consider a sequence of codes $(\varphi_{n},\Lset_{1n},\Lset_{2n},g_{1n}, g_{2n})$ such that the average probability of error tends to zero as $n \to \infty$.
This means, the probabilities $\prob{\hM_1 \neq M_1 | M_2}$, $\prob{\hM_2 \neq M_2 | M_1}$, and $\prob{(\hM_1,\hM_2) \neq (M_1,M_2)}$ are all bounded by some $\alpha_n$ that tends to zero as $n \to \infty$.
By Fano's inequality \cite{Cover_Thomas_2005}, it follows
\begin{align}
    H(M_1|\hM_1,M_2)\leq n\eps_{1n} \label{eq:fano1} \\
    H(M_2|M_1,\hM_2)\leq n\eps_{2n} \\
    H(M_1,M_2|\hM_1,\hM_2)\leq n\eps_{3n} \\
\end{align}
where $\eps_{kn}$ tend to zero as $n \to \infty$.

For $R_1$, we have
\begin{align}
    nR_1 &= H(M_1) \nonumber \\
    &= H(M_1|M_2) \nonumber \\
    &= I(M_1;\hM_1|M_2) + H(M_1|\hM_1,M_2) \nonumber \\
    &\leq I(M_1;\hM_1|M_2) + n\eps_{1n} \nonumber \\
    &\leq I(M_1;Y_1^n|M_2) + n\eps_{1n} \;, \label{eq:outerNR1}
\end{align}
where the first inequality follows from \eqref{eq:fano1} and the last one by the data processing inequality.
By similar arguments,
\begin{equation}
    nR_2 \leq I(M_2;Y_2^n|M_1) + n\eps_{2n} \label{eq:outerNR2}
\end{equation}
and
\begin{equation}
    n(R_1+R_2) \leq I(M_1,M_2;Y_1^n,Y_2^n) + n\eps_{3n} \,. \label{eq:outerNR12}
\end{equation}

Next, since we do not wish to have a regularized characterization, we need to further investigate the encoding POVMs $\Lset_{1n} \otimes \Lset_{2n}$ acting on $n$ consecutive channel uses.
Due to the same coding technique this is analogous to the proof in \cite[Section VIII]{Pereg_Deppe_Boche_2024}.
Consider \eqref{eq:outerNR1} again.
By the chain rule of mutual information, \eqref{eq:outerNR1} can be expressed as
\begin{equation}
    n(R_1 -\eps_{1n}) \leq \sum_{i=1}^n I\left(M_1;Y_1[i]|Y_1^{i-1},M_2\right) \label{eq:outerNR1chain}
\end{equation}
where $Y_1^n\equiv Y_1[1],...,Y_1[n]$, $Y^{i-1}\equiv Y_1[1],...,Y_1[i-1]$ for $i\in[2,n]$, and $Y^0\equiv\emptyset$.
For encoding, we perform the POVMs
\begin{equation}
    \Lset_1^{(m_1)} \otimes \Lset_2^{(m_2)} \;,
\end{equation}
where the encoding measurement at transmitter $k$, $k\in\{1,2\}$, depends solely on $m_k$.
We can write the measurement channel at transmitter 1 as
\begin{equation}
    \Tilde{\Lset}_{E_1\to X_1}^{(i,m_1)}(\psi) = \sum_{a_1\in\Xset_1}\left[\sum_{x_1^n\in\Xset_1^n:x_1[i]=a_1}\Tr{L^{(m_1)}_{x_1^n}\psi}\right]\ketbra{a_1}
\end{equation}
for $i\in[1,n]$ where the operator $L^{(m_1)}_{x_1^n}$ corresponds to a measurement acting on the quantum state $\psi\in\Dset(\Hset_{E_1})$ and produces a codeword of length $n$, while the above channel yields only the symbol at position $i$, $i\in[1,n]$.
Similarly, for encoder 2 we obtain
\begin{equation}
    \Tilde{\Lset}_{E_2\to X_2}^{(i,m_2)}(\psi) = \sum_{a_2\in\Xset_2}\left[\sum_{x_2^n\in\Xset_2^n:x_2[i]=a_2}\Tr{L^{(m_2)}_{x_2^n}\psi}\right]\ketbra{a_2}\,.
\end{equation}

Let $\varphi_{E_1E_2}=\sum_{j,l}\phi_{E_1,j}\otimes \phi_{E_2,l}$ be an arbitrary decomposition of the shared state.
By the linearity of the states and encoding operations,
\begin{align}
    (\Tilde{\Lset}_{E_1\to X_1}^{(i,m_1)}\otimes\Tilde{\Lset}_{E_1\to X_1}^{(i,m_1)})(\varphi_{E_1E_2}) &= \sum_{j,l}(\Tilde{\Lset}_{E_1\to X_1}^{(i,m_1)}\otimes\Tilde{\Lset}_{E_1\to X_1}^{(i,m_1)})(\phi_{E_1,j}\otimes\phi_{E_2,l}) \\
    &= \sum_{(a_1,a_2)\in\Xset_1\times\Xset_2}\left[\sum_{\substack{(x_1^n,x_2^n)\in\Xset_1^n\times\Xset_2^n:\\x_1[i]=a_1, x_2[i]=a_2}} \Tr{\left(L^{(m_1)}_{x_1^n}\otimes L^{(m_2)}_{x_2^n}\right)\varphi_{E_1E_2}}\right]\ketbra{a_1, a_2}\,.
\end{align}
Thus, the encoding operations for the instance $i$ are described based on the applied POVMs.
We deduce that the channel inputs $X_k[i]$ can be obtained from a product of measurements of form $L_k(x_k|i,V_0[i], V_k[i])$ for $k\in\{1,2\}$, where we define $V_0[i]\equiv (Y_1^{i-1},Y_2^{i-1})$, $V_k[i]\equiv (M_k,Y_1^{i-1},Y_2^{i-1})$ for $i\in[1,n]$.

\parbreak We can rewrite \eqref{eq:outerNR1chain} as
\begin{align}
    R_1-\eps_{1n}&\leq\frac{1}{n}\sum_{i=1}^n I(V_1[i];Y_1[i]|V_0[i],V_2[i]) \nonumber \\
    &=I(V_1[J];Y_1[J]|V_0[J],V_2[J],J) \,,
\end{align}
where the index $J$ determines the length of memory and is drawn uniformly at random from $\{1,...,n\}$.
Similarly, we obtain from \eqref{eq:outerNR2} and \eqref{eq:outerNR12}
\begin{align}
    R_1-\eps_{2n}&\leq I(V_2[J];Y_2[J]|V_0[J],V_1[J],J) \;, \\
    R_1+R_2-\eps_{3n}&\leq I(V_1[J],V_2[J];Y_1[J],Y_2[J]|V_0[J],J) \;.
\end{align}
The proof follows by setting $V_0 \equiv (V_0[J],J)$, $V_k \equiv V_k[J]$ and $Y_k \equiv Y_k[J]$ for $k\in\{1,2\}$.
\qed

\section{Cardinality Proof}
\label{sec:proofCardinality}

We bound the cardinality of the alphabet $\Vset_0$ for the time-sharing variable $V_0$.
Our goal is to demonstrate that the rate region $\mathcal{R}_{\text{ET-HK}}(P_{Y_1,Y_2|X_1,X_2})$ remains unchanged under a restriction of this alphabet.
We fix, without loss of generality, some input distribution $p_{V_0}p_{V_1,U_1|V_0}p_{V_2,U_2|V_0}$, a shared quantum state $\varphi_{E_1E_2}$ and encoding POVMs $\Lset_1 \otimes \Lset_2$ for the IC $P_{Y_1,Y_2|X_1,X_2}$.
Define the map $f: \Vset_0 \to \mathbb{R}^7$ as
\begin{align}
f(v_0)= \Big(
    &I(V_1,U_1;Y_1|V_0=v_0,U_2), \nonumber \\
    &I(V_2,U_2;Y_2|V_0=v_0,U_1), \nonumber \\
    &I(V_1;Y_1|V_0=v_0,U_1,U_2)+I(U_1,V_2,U_2;Y_2|V_0=v_0), \nonumber \\
    &I(V_1,U_2;Y_1|V_0=v_0,U_1)+I(U_1,V_2;Y_2|V_0=v_0,U_2), \nonumber \\
    &I(V_1,U_1,U_2;Y_1|V_0=v_0)+I(V_2;Y_2|V_0=v_0,U_1,U_2), \nonumber \\
    &I(V_1,U_1,U_2;Y_1|V_0=v_0)+I(V_1;Y_1|V_0=v_0,U_1,U_2)+I(U_1,V_2;Y_2|V_0=v_0,U_2), \nonumber \\
    &I(V_1,U_2;Y_1|V_0=v_0,U_1)+I(U_1,V_2,U_2;Y_2|V_0=v_0)+I(V_2;Y_2|V_0=v_0,U_1,U_2)
   \Big) \,. \label{eq:Cardnality_f}
\end{align}
We extend this map to operate on probability distributions $p_{V_0}$ in the following manner:
\begin{align} %
F: p_{V_0} \to \sum_{v_0 \in \Vset_0} p_{V_0}(v_0)f(v_0) = \Big(
    &I(V_1,U_1;Y_1|V_0,U_2), \nonumber \\
    &I(V_2,U_2;Y_2|V_0,U_1), \nonumber \\
    &I(V_1;Y_1|V_0,U_1,U_2)+I(U_1,V_2,U_2;Y_2|V_0), \nonumber \\
    &I(V_1,U_2;Y_1|V_0,U_1)+I(U_1,V_2;Y_2|V_0,U_2), \nonumber \\
    &I(V_1,U_1,U_2;Y_1|V_0)+I(V_2;Y_2|V_0,U_1,U_2), \nonumber \\
    &I(V_1,U_1,U_2;Y_1|V_0)+I(V_1;Y_1|V_0,U_1,U_2)+I(U_1,V_2;Y_2|V_0,U_2), \nonumber \\
    &I(V_1,U_2;Y_1|V_0,U_1)+I(U_1,V_2,U_2;Y_2|V_0)+I(V_2;Y_2|V_0,U_1,U_2)
   \Big) \,. \label{eq:Cardnality_F}
\end{align} %
Clearly, $F$ is a convex combination of points $f(v_0)$ in $\mathbb{R}^7$.
Therefore, it maps the set of distributions on $\Vset_0$ to a connected compact set in $\mathbb{R}^7$.
According to the Fenchel-Eggleston-Carathéodory theorem \cite{Eggleston_1966}, we know that any point in the convex closure of a connected compact set within $\mathbb{R}^d$ belongs to the convex hull of at most $d$ points from that set.
Thus, for every distribution $p_{V_0}$, there exists a probability mass function $\Bar{p}_{V_0}$ that is a convex combination of seven distributions $p_{V_0,i}$, $i\in[1,7]$, on $\Vset_0$, such that $F(\Bar{p}_{V_0})=F(p_{V_0})$.
By defining $f(v_{0,i}) \equiv F(p_{V_0,i})$, $i\in[1,7]$, we conclude that the alphabet size for the time sharing variable can be restricted to $|\Vset_0|\leq7$ while preserving the expressions in \eqref{eq:Cardnality_F} that appear in the bounds in \eqref{eq:HK_ineq1} -- \eqref{eq:HK_ineq1} on the achievable rate region.
Consequently, the achievable rate region remains unchanged.

\section{Conclusion and Discussion}
\label{sec:summaryDiscussion}

In this study, we have investigated communication over a two-sender, two-receiver classical interference channel with access to entanglement resources between the transmitters.
Our primary contribution is the establishment of an inner and an outer bound to the capacity region for a general IC with entangled transmitters.
Currently, there are no established cardinality bounds for auxiliary variables in this rate region other than the time-sharing variable, which remains a crucial aspect for future research.

\parbreak We have addressed the persistent challenge posed by the absence of a general capacity expression, even in the purely classical case, particularly regarding the evaluation of potential quantum advantages.
We have underlined the striking similarities among the inner bound expressions, which complicate the question of whether a quantum advantage truly exists.

\parbreak Nevertheless, we have demonstrated by an example that entanglement can indeed significantly enhance performance in certain channels.
Even though the example - based on the magic square game - is a very specific channel, the simplicity of this game is widely appreciated.
We see its application in various contexts to show a quantum advantage, including its potential relevance to device-independent quantum key distribution in communication networks \cite{Jain_Miller_Shi_2020}\cite{Zhen_Mao_Zhang_Xu_Sanders_2023}.
Other channel examples besides channels derived from non-local games that exhibit advantages from shared entanglement remain a topic of ongoing research.

\parbreak Our results on a more complex model of an interference channel including two transmitters and two receivers represent an advancement in the analysis of communication networks that incorporate quantum resources and entanglement.

\parbreak However, this work also brings forth further challenges.
We discussed that even under (very) strong interference conditions - where the capacity regions are well-understood in the classical case - the introduction of entanglement complicates our ability to apply similar arguments to establish tight upper bounds.
To address this issue, further investigation is required to understand how the entanglement and encoding POVMs influence channel inputs.
Additionally, to assess more realistic scenarios, we have to study the effects of noisy and unreliable entanglement resources on the observed advantage.

\appendix[Purification]
\label{app:purification}

The achievable rate region in \eqref{eq:Etx-HK-region} with entangled transmitters can be attained by considering only pure states in the union.
First, we consider a given Hilbert space $\mathcal{H}_{E_1} \otimes \mathcal{H}_{E_2}$ and demonstrate that the union over bipartite states $\varphi_{E_1E_2}$ can be exhausted by pure states.
Let us consider a specific distribution $p_{V_1,U_1|V_0}(\cdot|v_0) p_{V_2,U_2|V_0}(\cdot|v_0)$, an arbitrary bipartite state $\varphi_{E_1E_2}$ and measurements $\mathcal{L}_1(v_0,u_1,v_1)$ and $\mathcal{L}_2(v_0,u_2,v_2)$ on $E_1$ and $E_2$, respectively.
Let $\mathfrak{R}(\varphi,\Lset_1,\Lset_2)$ denote the associated rate region.
Consider the spectral decomposition
\begin{equation} \label{eq:Cardinality_Spectral}
    \varphi_{E_1E_2}=\sum_{z\in\Zset}p_Z(z)\ketbra{\phi_z}_{E_1E_2}
\end{equation}
of the joint state, where $p_Z$ is a probability distribution such that the pure states $\ket{\phi_z}_{A_1A_2}, z \in \Zset$ form an orthonormal basis for $\mathcal{H}_{E_1} \otimes \mathcal{H}_{E_2}$.
Then, there exists the purification
\begin{equation}
    \ket{\psi_{E_1E_2S_1S_2}}=\sum_{z \in \Zset}\sqrt{p_Z(z)}\ket{\phi_z}_{E_1E_2} \otimes \ket{z}_{S_1} \otimes \ket{z}_{S_2} \,.
\end{equation}
Now, we consider the rate region $\mathfrak{R}(\psi,\Lset'_1,\Lset'_2)$ that is associated with the following choice of state, distribution and measurements.
We set the distribution as before $p_{V_1,U_1|V_0}(\cdot|v_0) p_{V_2,U_2|V_0}(\cdot|v_0)$.
Let Alice 1 and Alice 2 share the state $\ket{\psi_{E_1S_1E_2S_2}}$, and suppose that Alice 1 performs a measurement on $(E_1, S_1)$ while Alice 2 performs a measurement on $(E_2,S_2)$ using the POVMs
\begin{align}
    L'_1(x_1,z|v_0,u_1,v_1) &= L(x_1|v_0,u_1,v_1) \otimes \ketbra{z}_{S_1} \,, \\
    L'_2(x_2,z|v_0,u_2,v_2) &= L(x_2|v_0,u_2,v_2) \otimes \ketbra{z}_{S_2} \,,
\end{align}
for $(v_0,u_1,v_1,u_2,v_2,x_1,x_2,z) \in \Vset_0 \times \Uset_1 \times \Vset_1 \times \Uset_2 \times \Vset_2 \times \Xset_1 \times \Xset_2 \times \Zset$.
The corresponding input distribution $p'_{X_1,X_2|V_0,U_1,V_1,U_2,V_2}$ is given by
\begin{align}
p'_{X_1,X_2|V_0,U_1,V_1,U_2,V_2}(x_1,x_2|v_0,u_1,v_1,u_2,v_2)&=
\sum_{z\in\Zset} p_Z(z) \trace{\left( L_1(x_1|v_0,u_1,v_1)\otimes L_2(x_2|v_0,u_2,v_2) \right) \ketbra{ \phi_z }_{E_1E_2}}
\\
&=
 \trace{\left( L_1(x_1|v_0,u_1,v_1)\otimes L_2(x_2|v_0,u_2,v_2) \right)\left(\sum_{z\in\Zset}  p_Z(z)\ketbra{ \phi_z }_{E_1 E_2} \right) }
\\
&=
 \trace{ \left( L_1(x_1|v_0,u_1,v_1)\otimes L_2(x_2|v_0,u_2,v_2) \right)\varphi_{E_1 E_2}}
\\
&=
p_{X_1,X_2|V_0,U_1,V_1,U_2,V_2}(x_1,x_2|v_0,u_1,v_1,u_2,v_2)
\end{align}
where the third equality follows from (\ref{eq:Cardinality_Spectral}), and the last one from \eqref{eq:Distribution_HK_inner}.
Consequently, we have $\mathfrak{R}(\psi,\Lset_1',\Lset_2')=\mathfrak{R}(\varphi,\Lset_1,\Lset_2)$.
From this, we deduce that the entire region $\mathcal{R}_{\text{ET-HK}}(P_{Y_1,Y_2|X_1,X_2})$ can be attained using only pure states.

\bibliographystyle{IEEEtran}
\bibliography{bibliography.bib}

% Generated by IEEEtran.bst, version: 1.14 (2015/08/26)
\begin{thebibliography}{10}
\providecommand{\url}[1]{#1}
\csname url@samestyle\endcsname
\providecommand{\newblock}{\relax}
\providecommand{\bibinfo}[2]{#2}
\providecommand{\BIBentrySTDinterwordspacing}{\spaceskip=0pt\relax}
\providecommand{\BIBentryALTinterwordstretchfactor}{4}
\providecommand{\BIBentryALTinterwordspacing}{\spaceskip=\fontdimen2\font plus
\BIBentryALTinterwordstretchfactor\fontdimen3\font minus \fontdimen4\font\relax}
\providecommand{\BIBforeignlanguage}[2]{{%
\expandafter\ifx\csname l@#1\endcsname\relax
\typeout{** WARNING: IEEEtran.bst: No hyphenation pattern has been}%
\typeout{** loaded for the language `#1'. Using the pattern for}%
\typeout{** the default language instead.}%
\else
\language=\csname l@#1\endcsname
\fi
#2}}
\providecommand{\BIBdecl}{\relax}
\BIBdecl

\bibitem{Granelli_Bassoli_Nötzel_Fitzek_Boche_daFonseca_Guerrieri_2022}
F.~Granelli, R.~Bassoli, J.~Nötzel, F.~H.~P. Fitzek, H.~Boche, N.~L.~S. da~Fonseca, and A.~Guerrieri, ``A novel architecture for future classical-quantum communication networks,'' \emph{Wireless Communications \& Mobile Computing}, vol. 2022, Jan. 2022.

\bibitem{Ekert_1991}
A.~K. Ekert, ``Quantum cryptography based on {Bell}’s theorem,'' \emph{Physical Review Letters}, vol.~67, no.~6, p. 661–663, Aug. 1991.

\bibitem{Bennett_Shor_Smolin_Thapliyal_1999}
C.~H. Bennett, P.~W. Shor, J.~A. Smolin, and A.~V. Thapliyal, ``Entanglement-assisted classical capacity of noisy quantum channels,'' \emph{Physical Review Letters}, vol.~83, no.~15, p. 3081–3084, Oct. 1999.

\bibitem{Hsieh_Devetak_Winter_2008}
M.-H. Hsieh, I.~Devetak, and A.~Winter, ``Entanglement-assisted capacity of quantum multiple-access channels,'' \emph{IEEE Transactions on Information Theory}, vol.~54, no.~7, p. 3078–3090, Jul. 2008.

\bibitem{Leditzky_Alhejji_Levin_Smith_2020}
F.~Leditzky, M.~A. Alhejji, J.~Levin, and G.~Smith, ``\BIBforeignlanguage{en}{Playing games with multiple access channels},'' \emph{\BIBforeignlanguage{en}{Nature Communications}}, vol.~11, no.~1, p. 1497, Mar. 2020.

\bibitem{Nötzel_2020}
J.~Nötzel, ``Entanglement-enabled communication,'' \emph{IEEE Journal on Selected Areas in Information Theory}, vol.~1, no.~2, p. 401–415, Aug. 2020.

\bibitem{Seshadri_Leditzky_Siddhu_Smith_2023}
A.~Seshadri, F.~Leditzky, V.~Siddhu, and G.~Smith, ``On the separation of correlation-assisted sum capacities of multiple access channels,'' \emph{IEEE Transactions on Information Theory}, vol.~69, no.~9, p. 5805–5844, Sep. 2023.

\bibitem{Pereg_Deppe_Boche_2023}
U.~Pereg, C.~Deppe, and H.~Boche, ``The multiple-access channel with entangled transmitters,'' \emph{IEEE Global Communications Conference}, p. 3173–3178, Dec. 2023.

\bibitem{Yard_Hayden_Devetak_2011}
J.~Yard, P.~Hayden, and I.~Devetak, ``Quantum broadcast channels,'' \emph{IEEE Transactions on Information Theory}, vol.~57, no.~10, p. 7147–7162, Oct. 2011.

\bibitem{Pereg_Deppe_Boche_2021}
U.~Pereg, C.~Deppe, and H.~Boche, ``Quantum broadcast channels with cooperating decoders: an information-theoretic perspective on quantum repeaters,'' \emph{IEEE International Symposium on Information Theory}, p. 772–777, Jul. 2021.

\bibitem{Prabhakaran_Viswanath_2011}
V.~M. Prabhakaran and P.~Viswanath, ``Interference channels with source cooperation,'' \emph{IEEE Transactions on Information Theory}, vol.~57, no.~1, p. 156–186, Jan. 2011.

\bibitem{Quek_Shor_2017}
Y.~Quek and P.~W. Shor, ``Quantum and superquantum enhancements to two-sender, two-receiver channels,'' \emph{Physical Review A}, vol.~95, no.~5, p. 052329, May 2017.

\bibitem{Clauser_Horne_Shimony_Holt_1969}
J.~F. Clauser, M.~A. Horne, A.~Shimony, and R.~A. Holt, ``Proposed experiment to test local hidden-variable theories,'' \emph{Physical Review Letters}, vol.~23, no.~15, p. 880–884, Oct. 1969.

\bibitem{Popescu_Rohrlich_1994}
S.~Popescu and D.~Rohrlich, ``\BIBforeignlanguage{en}{Quantum nonlocality as an axiom},'' \emph{\BIBforeignlanguage{en}{Foundations of Physics}}, vol.~24, no.~3, p. 379–385, Mar. 1994.

\bibitem{Brassard_Broadbent_Tapp_2005}
G.~Brassard, A.~Broadbent, and A.~Tapp, ``Quantum pseudo-telepathy,'' \emph{Foundations of Physics}, vol.~35, no.~11, p. 1877–1907, Nov. 2005.

\bibitem{ElGamal_Kim_2011}
A.~El~Gamal and Y.-H. Kim, \emph{Network information theory}.\hskip 1em plus 0.5em minus 0.4em\relax Cambridge: Cambridge University Press, 2011.

\bibitem{Han_Kobayashi_1981}
T.~Han and K.~Kobayashi, ``A new achievable rate region for the interference channel,'' \emph{IEEE Transactions on Information Theory}, vol.~27, no.~1, p. 49–60, Jan. 1981.

\bibitem{Chong_Motani_Garg_ElGamal_2008}
H.-F. Chong, M.~Motani, H.~K. Garg, and H.~El~Gamal, ``On the {Han}–{Kobayashi} region for the interference channel,'' \emph{IEEE Transactions on Information Theory}, vol.~54, no.~7, p. 3188–3195, Jul. 2008.

\bibitem{Nair_Xia_Yazdanpanah_2015}
C.~Nair, L.~Xia, and M.~Yazdanpanah, ``Sub-optimality of {Han}-{Kobayashi} achievable region for interference channels,'' \emph{IEEE International Symposium on Information Theory}, p. 2416–2420, Jun. 2015.

\bibitem{Sato_1977}
H.~Sato, ``Two-user communication channels,'' \emph{IEEE Transactions on Information Theory}, vol.~23, no.~3, p. 295–304, May 1977.

\bibitem{Pereg_Deppe_Boche_2024}
U.~Pereg, C.~Deppe, and H.~Boche, ``The multiple-access channel with entangled transmitters,'' Feb. 2024, arXiv:2303.10456 [quant-ph].

\bibitem{Eggleston_1966}
H.~G. Eggleston, ``\BIBforeignlanguage{en}{Convexity},'' \emph{\BIBforeignlanguage{en}{Journal of the London Mathematical Society}}, vol. s1-41, no.~1, p. 183–186, 1966.

\bibitem{Costa_Gamal_1987}
M.~Costa and A.~Gamal, ``The capacity region of the discrete memoryless interference channel with strong interference (corresp.),'' \emph{IEEE Transactions on Information Theory}, vol.~33, no.~5, p. 710–711, Sep. 1987.

\bibitem{Csiszár_Körner_2011}
I.~Csiszár and J.~Körner, \emph{Information Theory: Coding Theorems for Discrete Memoryless Systems}, 2nd~ed.\hskip 1em plus 0.5em minus 0.4em\relax Cambridge: Cambridge University Press, 2011.

\bibitem{Kramer_2008}
G.~Kramer, ``\BIBforeignlanguage{en}{Topics in multi-user information theory},'' \emph{\BIBforeignlanguage{en}{Foundations and Trends® in Communications and Information Theory}}, vol.~4, no. 4–5, p. 265–444, 2008.

\bibitem{Cover_Thomas_2005}
T.~M. Cover and J.~A. Thomas, \emph{\BIBforeignlanguage{en}{Elements of Information Theory}}, 1st~ed.\hskip 1em plus 0.5em minus 0.4em\relax Hoboken, New Jersey: Wiley, 2005.

\bibitem{Jain_Miller_Shi_2020}
R.~Jain, C.~A. Miller, and Y.~Shi, ``Parallel device-independent quantum key distribution,'' \emph{IEEE Transactions on Information Theory}, vol.~66, no.~9, p. 5567–5584, Sep. 2020.

\bibitem{Zhen_Mao_Zhang_Xu_Sanders_2023}
Y.-Z. Zhen, Y.~Mao, Y.-Z. Zhang, F.~Xu, and B.~C. Sanders, ``Device-independent quantum key distribution based on the {Mermin}-{Peres} magic square game,'' \emph{Physical Review Letters}, vol. 131, no.~8, p. 080801, Aug. 2023.

\end{thebibliography}

\end{document}